# Observation of long-range ferromagnetism via anomalous supercurrents in a spin-orbit coupled superconductor

B. K. Xiang[1,*], Y. S. Lin[1,*], Q. S. He[1,*], J. J. Zhu[1], B. R. Chen[1], Y. F. Wang[1], K. Y. Liang[1], Z. J. Li[1], H. X. Yao[1], C. X. Wu[2], T. Y. Zhou[2], M. H. Fang[2], Y. Lu[3], I. V. Tokatly[4,5,6], F. S. Bergeret[3,4], Y. H. Wang[1,7,†]

1. State Key Laboratory of Surface Physics and Department of Physics, Fudan University, Shanghai 200433, China
2. School of Physics, Zhejiang University, Hangzhou 310058, China
3. Centro de Fisica de Materiales (CFM-MPC), Centro Mixto CSIC-UPV/EHU, Manuel de Lardizabal 4, E-20018 San Sebastian, Spain
4. Donostia International Physics Center (DIPC), Manuel de Lardizabal 5, E-20018 San Sebastian, Spain
5. Nano-Bio Spectroscopy group, Departamento Física de Materiales, Universidad del País Vasco, Av. Tolosa 72, E-20018 San Sebastián, Spain
6. IKERBASQUE, Basque Foundation for Science, E-48011 Bilbao, Spain
7. Shanghai Research Center for Quantum Sciences, Shanghai 201315, China

* These authors contributed equally to this work.
† To whom correspondence should be addressed. wangyhv@fudan.edu.cn

**Abstract**

**Conventional superconductors naturally disfavor ferromagnetism because the supercurrent-carrying electrons are paired into anti-parallel spin singlets. In superconductors with strong Rashba spin-orbit coupling, impurity magnetic moments induce supercurrents through the spin-galvanic effect. As a result, long-range ferromagnetic interaction among the impurity moments may be mediated through such anomalous supercurrents in a similar fashion as in itinerant ferromagnets. Fe(Se,Te) is such a superconductor with topological surface bands, previously shown to exhibit quantum anomalous vortices around impurity spins. Here, we take advantage of the flux sensitivity of scanning superconducting quantum interference devices to investigate superconducting Fe(Se,Te) in the regime where supercurrents around impurities overlap. We find homogeneous remanent flux patterns after applying a supercurrent through the sample. The patterns are consistent with anomalous edge and bulk supercurrents generated by in-plane magnetization, which occur above a current threshold and**

**follow hysteresis loops reminiscent of those of a ferromagnet. Similar long-range magnetic orders can be generated by Meissner current under a small out-of-plane magnetic field. The magnetization weakens with increasing temperature and disappears after thermal cycling to above superconducting critical temperature; further suggesting superconductivity is central to establishing and maintaining the magnetic order. These observations demonstrate surface anomalous supercurrents as a mediator for ferromagnetism in a spin-orbit coupled superconductor, which may potentially be utilized for low-power cryogenic memory.**

**Introduction**

Charge carriers in a solid material are capable of mediating spin correlation between localized magnetic moments. Ruderman-Kittel-Kasuya-Yosida (RKKY) [1–3] interaction is an important mechanism to align localized impurity spins into long-range ferromagnetic order through exchange coupling with conduction electrons in normal metals. If a metal also has strong Rashba spin-orbit coupling (SOC), RKKY interaction may further facilitate ferromagnetism through an applied current [4,5]. In an insulator, however, the RKKY mechanism is suppressed due to the lack of low energy quasiparticles. A magnetically doped topological insulator is an example of such a system, which forms long-range ferromagnetism by the Van Vleck mechanism on its way to becoming a quantum anomalous Hall insulator [6,7].

SOC in *s*-wave superconductors may drive the coupling between the spin of magnetic impurities and supercurrent. In the bulk of such a superconductor with SOC, the local magnetic moment of the impurity spin $S$ is locked with the orbital moment of supercurrent around it to generate spontaneous vortices without a magnetic field, the so-called quantum anomalous vortices (Fig. 1a) [8,9]. The sparsely populated vortices favor anti-parallel out-of-plane alignment to reduce the free energy of the supercurrent. Vortex-antivortex pairs form as a result [8]. For superconductors without SOC, because of the superconducting gap, the RKKY mechanism between the impurity moments through quasiparticle excitations is exponentially suppressed [10] just like in an insulator. Interestingly, in superconductors with Rashba SOC, non-exponential decaying interaction between magnetic impurities is predicted from RKKY-type calculation of the spin-spin correlation function [11]. A complete analysis of such magnetic coupling, which includes the London-Pearl screening effects, shows that the interaction of the anomalous supercurrents stemming from different impurities can mediate a long-range ferromagnetic coupling [12]. (If the impurity density is too high, the superconducting gap will be completely suppressed,

and a ferromagnetic order may appear via the RKKY mechanism at the expense of superconductivity [10,13].)

In Rashba superconductors and in superconductors with surfaces supporting a Dirac band, the symmetry allows for a term in the Ginzburg-Landau free energy of the form of Lifshitz invariant $F_L = \kappa(\boldsymbol{n} \times \boldsymbol{B}) \cdot \hbar \nabla \varphi$ [14–16], where $\nabla \varphi$ is a phase gradient in the order parameter, $\boldsymbol{B}$ is the magnetic induction, $\boldsymbol{n}$ is the surface normal, and $\kappa$ is the Edelstein constant, which depends on the Rashba splitting of the Kramer's pair at the Fermi level. For Dirac surface states, it is a limiting case of a large Rashba SOC when the physics is dominated by only one helical band. A direct consequence of the Lifshitz invariant is that any supercurrent $\boldsymbol{J_S}$ (being proportional to $\nabla \varphi$) induces polarization of surface impurity moments, which has been observed on isolated quantum anomalous vortex and antivortex in Fe(Se,Te) [8].

The inverse effect, in analogy with the spin-galvanic effect in SOC metals, is that a local magnetic moment polarizes the condensate and generates anomalous supercurrents $\boldsymbol{J_a} \propto \boldsymbol{n} \times \boldsymbol{S}$ at the center of the impurity [11,17–19]. Surrounding the impurity, $\boldsymbol{J_a}$ follows a dipolar distribution that decays as a power law and extends over long distances away from the magnetic impurity (Fig. 1b) [12]. Confined to the surface plane, two impurity spins favor an alignment along their center line (Fig. 1c). Such a configuration minimizes the total supercurrent on both impurities and thus reduces the free energy. For disordered impurity spins on a plane, ferromagnetic coupling leads to long-range in-plane magnetization $\boldsymbol{M_J}$ which generates an edge $\boldsymbol{J_a}$ with an opposite flow direction from that in the bulk (Fig. 1d) [15]. With the exception of square lattices where anti-ferromagnetism is favored [12], Rashba SOC allows the interesting possibility that a ferromagnetic order in a system of magnetic impurities can be established via coupling to the superconducting condensate.

Just like in a ferromagnet, where an external magnetic field is necessary to align magnetic domains, a long-range ferromagnetic order across the sample needs to be facilitated by applying a global supercurrent. In magnetic heterostructures with Rashba SOC, the magnetization follows hysteretic switching under external current [5,20]. The magnetic effect of an external supercurrent on a Rashba or Dirac superconductor with magnetic impurities has yet to be experimentally studied. In general, in the absence of a constant bias current or external magnetic field, both magnetization and anomalous supercurrents produce magnetic induction, the spatial distribution of which depends on sample geometry. Measuring such a magnetic signal and its evolution after applying a current may provide key evidence for the supercurrent-mediated long-range ferromagnetic coupling of magnetic impurities.

The detection of current-induced magnetization in magnetic heterostructures typically relies on Hall resistance, which is regarded as proportional to the magnetization. Such a Hall response does not occur in a superconductor due to the vanishing resistance. Therefore, a technique directly sensitive to the magnetic induction from the magnetization and anomalous supercurrents is essential for detecting long-range magnetization in Rashba superconductors. In this work, we use scanning superconducting quantum interference device (sSQUID) microscopy [21–30] to perform magnetic imaging of a Rashba superconductor Fe(Se,Te) in the high interstitial Fe density regime.

**Observation of hysteretic magnetization with bias current in Fe(Se,Te)**
Fe-chalcogenide superconductor Fe(Se,Te) has exhibited strong spin-orbit coupling, which leads to superconducting topological surface states [31–35]. The excess Fe atoms occupy interstitial sites and act as magnetic impurities [36]. Scanning tunneling microscopy studies have found zero-energy bound states in vortex cores or interstitial Fe sites on its surface [33,37–39]. Signs of time-reversal symmetry breaking have also been reported on the surface states from photoemission spectroscopy [40] and nitrogen-vacancy center magnetometry [41]. However, these works have been carried out on Fe(Se,Te) samples in the low interstitial Fe density regime without applying any supercurrent.

A high interstitial Fe concentration will completely suppress superconductivity [42–44] but the impurity moments are not exchange-coupled if the density is too low. The impurity concentration we consider here have an average interstitial Fe distance close to the coherence length so that their critical temperatures are lower than the bulk superconducting samples [45,46]. For the particular sample shown in Figure 2a, its critical temperature is about 6 K (Fig. S1a). The suppression of superconductivity by magnetic impurities is more visible from the reduced diamagnetic susceptibility (Fig. 2b) compared with that of the samples with <1% of interstitial Fe [8]. Flake samples with excess interstitial Fe tend to exhibit inhomogeneity in superfluid density [47] and this is evident from separated domains with different diamagnetic strength (Fig. 2b, dashed regions). The critical current of the lower domain is around 250 μA (Fig. 2c) when applying the current through the two top current leads, as indicated in Figure 2a. And unlike the quantum anomalous vortices observed in those samples, we do not observe any isolated vortices here, regardless of the cooling field.

We image the flux response of Fe(Se,Te) under a bias current ($I_b$) at the base temperature of 1.5 K. There is no flux contrast after cooling the sample under zero-field (Fig. 2d). Under a direct current of $I_b = 100$ µA flowing from the top-left electrode to the one on the top-right, the flux signal is dominated by an out-of-plane Oersted field, which has opposite signs on the two sides of the current (Fig. 2e). The current distribution extracted from the flux image shows that the top part of the sample has a higher current density (Fig. 2f). As a result, the magnetic induction on most of the sample is less than 1 mG. Interestingly, the current flux within the sample changes sign when $I_b = 200$ µA (Fig. 2g). The sign change rules out Oersted field-generated vortices. The current distribution (Fig. 2h) shows a strong current density along the lower sample edge, which is consistent with earlier observations [42]. After $I_b = 200$ µA is removed (Fig. 2i), there is remanent flux with the same positive sign as the flux signal within the sample when the bias is applied (Fig. 2d). Reversing the direction of bias current to $I_b = -200$ µA, we obtain a current flux image that shows the opposite sign (Figs. 2j and k). Removing $I_b = -200$ µA also leaves remanent flux (Fig. 2l) but with the opposite sign from those after removing the positive peak current $I_p = 200$ µA (Fig. 2i).

We have observed such remanent flux in all the superconducting Fe(Se,Te) samples with a modest amount of Fe impurities (Fig. S7). But this effect is absent both in low interstitial Fe density samples in which quantum anomalous vortices are present and in highly doped non-superconducting samples. The induced magnetization does not affect diamagnetic susceptibility (Fig. S2b). The pattern of the remanent flux at $I_b = 0$ µA after $I_p = 200$ µA is applied is similar to the flux image at $I_b = 200$ µA after subtracting the flux contribution from the supercurrent ($\Phi(I_b = 200$ µA$) - 2 \times \Phi(I_b = 100$ µA$)$) (Fig. S2a). The memory effect suggests a long-range magnetization facilitated by the bias supercurrent above a certain threshold.

In order to understand how the remanent flux appears with the bias, we park the nano-SQUID over the lower-right domain (Fig. 2d, yellow circle) and sweep $I_b$ continuously. Any remanent magnetization from previous rounds is erased by raising the temperature above $T_c$ and cooling down under zero-field and zero $I_b$ again. The thermal cycling results in a virgin state with zero flux (Fig. 3a, empty circle). As $I_b$ is increased from zero, the local flux signal $\Phi_d$ is linearly proportional to $I_b$ with a negative slope. At $I_b \sim 140$ µA, $\Phi_d$ suddenly turns upwards and quickly crosses zero at $I_b \sim 160$ µA. At $I_b = 200$ µA, $\Phi_d$ reaches the level consistent with the image (Fig. 2c). Upon reducing $I_b$, $\Phi_d$ first reduces slightly and then increases such that $\Phi_d$ is positive when $I_b = 0$. It keeps increasing when $I_b$ changes sign until $I_b = -120$ µA, when $\Phi_d$ suddenly turns down sharply, crosses zero and then reaches negative value at $I_b = -200$ µA. The upsweep from $-200$ µA is antisymmetric with the down sweep from $200$ µA, and it merges with the virgin trace at $I_b = 140$ μA, the current at which $\Phi_d$ initially turned. Subsequent $I_b$ sweeps follow the large hysteretic loop. By subtracting the Oersted field contribution (which is linear with $I_b$) from the current flux, we obtain a hysteresis loop of induced flux purely from the polarization of the sample by the supercurrent (Fig. 3b). It is reminiscent of a magnetization switched by current-induced spin-orbit torque in heavy metal-ferromagnet heterostructures [5].

**Anomalous supercurrents and long-range magnetization**

Since we notice an enhanced current density along the sample edge under a high bias (Figs. 2h and k), we investigate how it evolves from a low bias regime and its relation with the magnetic hysteresis. We take line cuts across the edge (Fig. 2e, blue line) from the flux images obtained under increasing $I_b$ after a zero-field cooling (Fig. S3). The flux contrast increases very slowly for small $I_b$ but shows a discontinuously sharp enhancement above 120 μA (Fig. 3c). The supercurrent density at the edge similarly exhibits a jump at this current (Fig. 3d), which is the same as the switching current in the ferromagnetic hysteresis loop (Fig. 3b). Both the intensity and the sign of the flux from this edge current are consistent with the remanent flux (Figs. 2i and l). This provides clear evidence that an anomalous supercurrent on the edge occurs after the bias current exceeds a threshold and remains there after the bias is removed.

As discussed in the introduction, there are potentially two main sources that may contribute to the remanent flux signal in a Rashba superconductor. Besides the anomalous supercurrents $J_a$, the magnetization $M_J$ mediated by the supercurrent also produces magnetic induction (Figs. 1c and d). Since our sample has a finite thickness (about 300 nm), both the top and bottom surfaces have to be taken into account (Fig. 3e). Under an in-plane bias current $I_b$, supercurrent passes through both surfaces equally and points in the same direction. But since the surface normals $n$ are reversed on the two surfaces, their $M_J$, which are in-plane, are opposite from each other (Fig. 3e). Given that our scanning height of 1 μm is much larger than the separation between two surfaces, their $M_J$ flux signal basically cancels. On the other hand, since $J_a$'s from the two surfaces generated by $M_J$ go through another cross product with $n$, their signs are identical. Therefore, the remanent flux signal is twice the flux of $J_a$ from each surface, whose magnitude is proportional to its $M_J$.

Using the above vectorial relations between magnetization and supercurrent, we can now understand the hysteretic magnetic behavior of Fe(Se,Te) with current bias. Under a current bias applied through the sample, a unidirectional supercurrent $J_s$ is established (Fig. 2h, blue lines). Noting that $J_a$ has opposite directions in the bulk and along the edges (Figs. 1d and 3e), we can see $J_a$ in the bulk (Fig. 2h, dark green lines) counters $J_S$ from the external bias, while the edge $J_a$ (Fig. 2h, light green lines) flows in the same direction as $J_S$. Although the total anomalous currents carried by the bulk and edge are the same, the current density is much higher along the edge (Figs. 1c and d). This causes a much-enhanced edge current density (Figs. 2h, k and 3c) after the onset of $M_J$ when $I_b$ exceeds the threshold of 120 µA. Once the bias is removed, $J_S$ disappears and there is no net current through the sample (Fig. 2i). The edge and bulk $J_a$ circles around the sample with a distribution determined by the pattern of remanent $M_J$, which depends on how $I_b$ is applied and the geometry of the sample (Fig. S4).

The ferromagnetic coupling mediated by the anomalous supercurrents between impurity moments is particularly sensitive to the loss of superconducting condensate. To show this, we again first generate the remanent flux by applying and then removing $I_b = -200$ µA at the base temperature. We then increase the sample to a specific temperature $T_a$ and then cool it back to the base temperature for imaging (Fig. S5). The flux contrast reduces with increasing $T_a$ and disappears above 5 K. Plotting the flux contrast with resistance as a function of temperature (Fig. 3f), we find that $M_J$ is roughly constant below $T_a = 3$ K and starts to decrease appreciably above that. While the superconductor is still in a zero-resistance state at 4 K, which is also evident from a small but finite superfluid density on the right domain of the sample (Fig. 3f), $M_J$ is halved upon reaching this temperature. At 5 K, the resistance is still an order of magnitude smaller than the normal state resistance. And yet, $M_J$ is almost completely invisible. This suggests that the long-range ferromagnetic ordering between magnetic impurities is disrupted by thermal fluctuations in the underlying superconducting condensate.

**Magnetization with an external field**
The ferromagnetic coupling between impurity moments can not only be mediated by a bias current but also by a circulating supercurrent from the Meissner effect. After zero-field cooling (Fig. 4a), we apply an out-of-plane magnetic field $H$ at base temperature much smaller than the lower critical field of Fe(Se,Te) (~ 500 Oe) [8] to induce the Meissner current. We image the flux signal with the field present and change the field after each scan. The flux response is diamagnetic at $H = 0.64$ G for both the top and bottom domains (Fig. 4a). But at $H = 1.08$ G the bottom part starts to turn positive, and the paramagnetic area grows bigger and stronger with increasing field (Fig. S6). At $H = 1.51$ G, the lower domain is completely paramagnetic, while the top domain is still diamagnetic. Upon reducing $H = 0.64$ G, however, the top domain also changes sign so that both domains are paramagnetic. This is in clear contrast with the flux signal at the same field from the upsweep. The positive flux signal remains after removing the field. The flux pattern after reaching $H =$ -1.51 G in a down-sweep and then returning to -0.64 G in an up-sweep is anti-symmetric with that at 1.51 G and 0.64 G in a reversed field history, respectively. And the remanent flux signal is negative upon returning to zero. The whole hysteresis loop with $H$ is better visualized by sampling one location within each domain (Fig. 4b). The overall hysteresis is consistent with the $M_J$ induced by a bias supercurrent.

The remanent flux contrast induced by the Meissner current appears similar to that induced by the bias current. This is the case only because of the specific current leads used in the bias experiment. When we use electrodes located on the same side of the sample, a large part of the supercurrent circles around the sample (Figs. 2h and Figs. S4a-b), which is similar to the current distribution of the Meissner current (Fig. 4a). As a result, the magnetization texture is also composed of in-plane moments pointing towards or away from the center of the sample. However, when we use electrodes on the opposite side of the sample, the supercurrent flows unidirectionally (Fig. 3e and Fig. S4c). The remanent flux we obtain in this scenario is consistent with a uniform in-plane magnetization. A more thorough analysis of this scenario to examine the vectorial product relation between $M_J$ and $J_S$ due to the magneto-electric effect on the superconducting Dirac surface states is being carried out in a separate work.

**Discussion and outlook**

Recent theory [12] predicts ferromagnetic coupling between two impurity moments as long as they are in-plane, even if their distance is larger than $\lambda$. But the ferromagnetic coupling energy decays with the distance, and it has to compete with the out-of-plane polarization in the quantum anomalous vortex-antivortex state, which is favored for isolated vortices [8,9]. A spontaneously tilted spin orientation [48] with antiferromagnetic out-of-plane and ferromagnetic in-plane configuration may occur at a distance larger than $\lambda$ (Fig. 1b). The ferromagnetic order is also predicted over the antiferromagnetic order for magnetic impurities on a 2D oblique lattice in a large parameter space [12]. However, a full calculation for a disordered impurity distribution such as that in our experimental system is too computationally involved to perform at this point. We can only speculate here that the ferromagnetic order may be favored since the antiferromagnetic order typically requires a regular lattice without frustration, while ferromagnetic order depends less on the lattice [49]. Although a disordered state is also possible under three-dimensional random-site dipole interactions [50], the extra ferromagnetic term besides the dipole interaction in a two-dimensional Rashba superconductor [12] may stabilize a ferromagnetic order.

At last, we compare ferromagnetism in Rashba superconductors and conventional magnetic heterostructures with Rashba-type SOC with regard to potential applications. In both cases, the basic mechanism for switching is current-induced spin-orbit-torques [5]. Therefore, the magnetic dynamics similarly follow the Landau-Lifshitz-Gilbert equation, which determines the fundamental timescale of the switching process. In the ferromagnet/heavy metal heterostructures, the switching current density is typically on the order of $10^6$ A/cm$^2$. In our Fe(Se,Te) superconductor, the switching current of 120 µA over a 50-µm-wide 300-nm-thick cross-section corresponds to a current density less than $1 \times 10^3$ A/cm$^2$. Since only the surface supercurrent is essential for the switching, it is plausible to further reduce the switching current density by fabricating thin film devices in order to reduce the current shunted by the bulk. Another significant advantage of being in the superconducting state while switching the magnetization is the elimination of power dissipation. One disadvantage of using a superconductor is that the magnetic state cannot be directly read out through the Hall resistance of the device. Nevertheless, electrical detection of the memory status is not difficult. Besides using SQUID as a flux-to-voltage converter, as we demonstrate, the anomalous supercurrents may be detected by the phase difference across Josephson junctions [51] in contact with the surface of Fe(Se,Te). In terms of non-volatility in the event that a low-temperature environment is lost, it is conceivable that the superconducting Dirac surface state may be proximitized by a conventional ferromagnet with easy-plane anisotropy as an interface between cryogenic and room-temperature spintronics [52,53].

In conclusion, we use sSQUID microscopy to uncover the anomalous supercurrent-mediated ferromagnetic coupling between magnetic impurity moments in Fe(Se,Te). Supercurrent, whether it is applied through external bias or generated through the Meissner effect under a small out-of-plane magnetic field, polarizes impurity moments to form long-range surface magnetization. An anomalous edge supercurrent appears above a bias current threshold and remains after the bias is removed, which manifests the interaction mechanism. The temperature dependence of the remanent magnetization shows that the superconducting condensate is essential for the ferromagnetic interaction. These experimental results support a general mechanism in superconductors with Rashba SOC or Dirac surface states to mediate ferromagnetic coupling between localized magnetic moments through the condensate. The switchable and non-volatile magnetization state may serve as a cryogenic memory that can be interfaced with ferromagnets to facilitate the dissipationless conversion between spin and charge currents for applications in superconducting spintronics.

## Acknowledgement

We would like to acknowledge support by National Key R&D Program of China (Grant No. 2021YFA1400100), National Natural Science Foundation of China (Grant No. 11827805 and 12150003) and Shanghai Municipal Science and Technology Major Project (Grant No. 2019SHZDZX01). Work at Zhejiang University is supported by National Key R&D program of China (Grant No. 2022YFA1403202) and the National Natural Science Foundation of China (Grant No. 12074335). F.S.B and Y.L. acknowledge financial support from Spanish MCIN/AEI/ 10.13039/501100011033 through project PID2020-114252GB-I00 (SPIRIT) and TED2021-130292B-C42, and the Basque Government through grant IT-1591-22. I.V.T. acknowledges support by Grupos Consolidados UPV/EHU del Gobierno Vasco (Grant IT1453-22) and by the grant PID2020-112811GB-I00 funded by MCIN/AEI/10.13039/501100011033. All the authors are grateful for the discussion with S. Y. Yin, Y. Z. Wu and X. P. Qiu and experimental assistance by Q. He, L. Zhou and C. L. Zheng.

# Figures

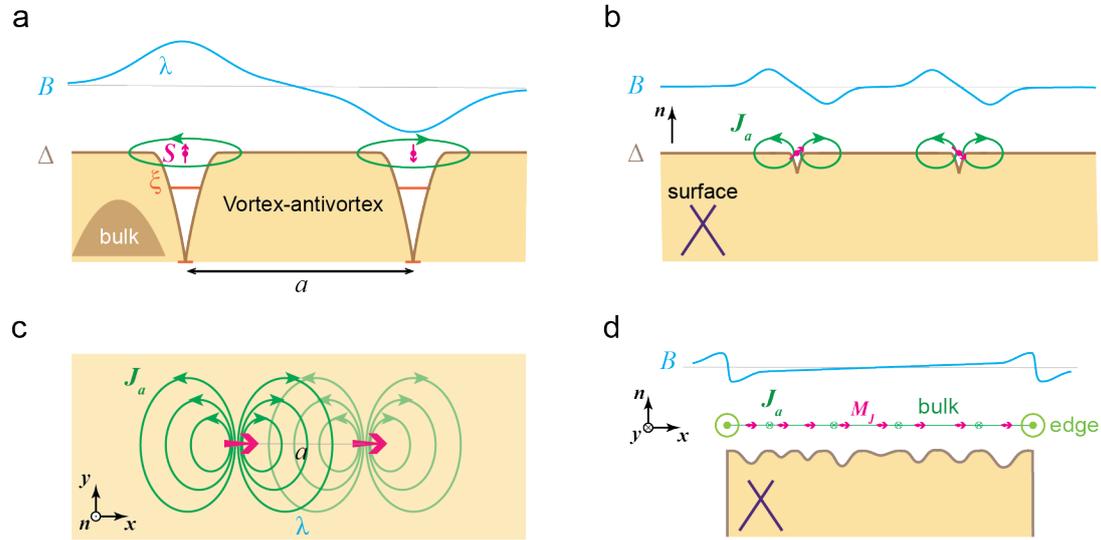

**Figure 1. Ferromagnetic coupling between impurity moments via anomalous supercurrents on a Dirac surface band. a**, For bulk states in a spin-orbit coupled superconductor, when two impurity distance $a$ is large (relative to the London penetration depth $\lambda$ of the superconductor), the impurity moments $S$ (purple arrows) point out-of-plane to couple with a circulating supercurrent (green circles) and generate quantum anomalous vortices. **b**, For two impurity moments on the surface Dirac band, the in-plane component of $S$ generates anomalous supercurrents $J_a$ (green lines) due to the surface Rashba field following the cross product: $J_a \propto n \times S$, where $n$ is the surface normal. The distribution of $J_a$ around each spin has a dipole-like pattern. **c**, Top view of two impurity moments on the surface interacting through the anomalous supercurrents. When their distance $a \lesssim \lambda$, the spins align along their center line to minimize the free energy. **d**, Cross-sectional view of impurity moments on a finite surface. The ferromagnetic coupling among impurities to form an in-plane long-range magnetization $M_J$ may be favored (with the exception of a square lattice). The $J_a$ in the bulk has a smaller magnitude and flows in the opposite direction from the $J_a$ on the edge.

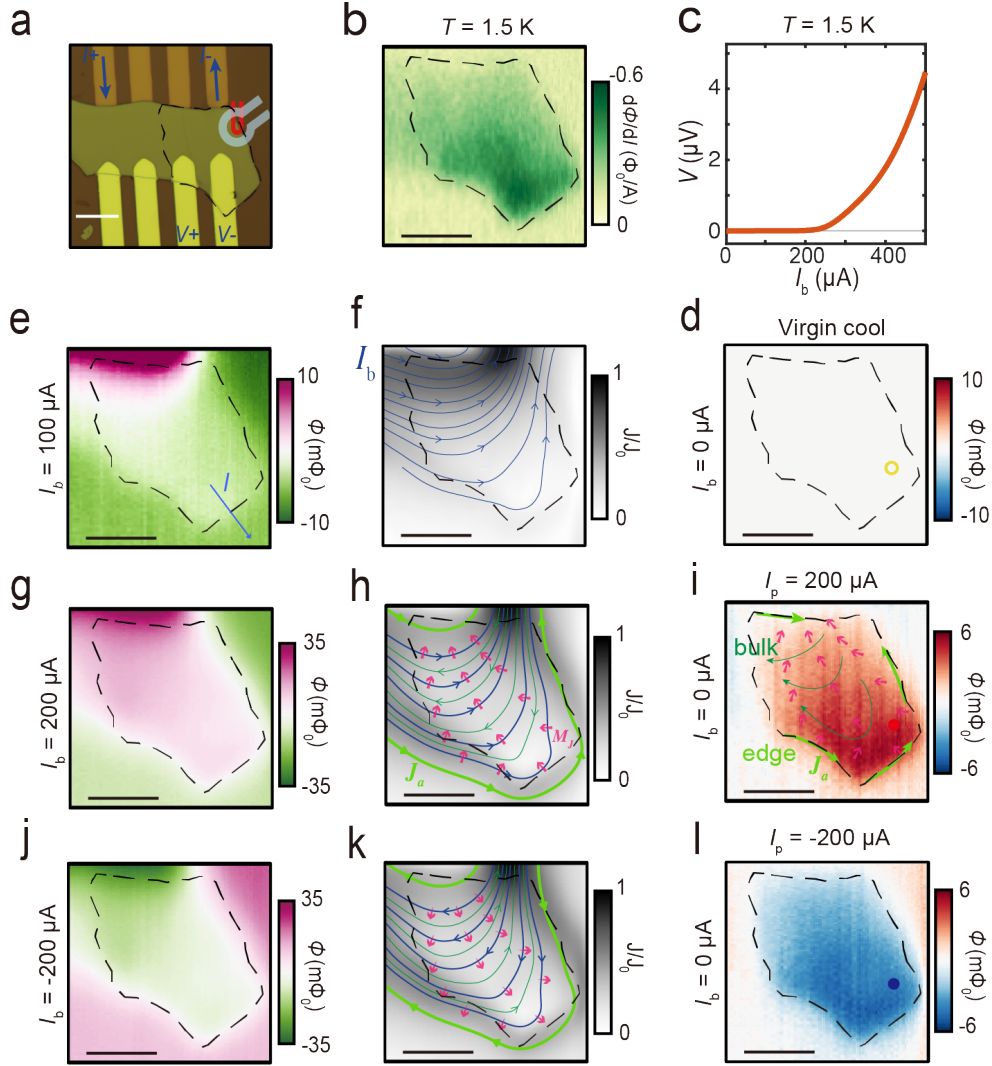

**Figure 2. Supercurrent-induced magnetization. a**, Optical image of the Fe(Se,Te) sample. Here, we use the outer two electrodes on the upper edge (blue arrows) for the source and drain of a bias current $I_b$. **b**, Susceptometry image of right half of the sample at $T = 1.5$ K. The lower part shows stronger diamagnetism than the upper part. **c**, Current-voltage characteristics of the lower part with the voltage drop measured across the probes shown in **a**. The critical current of this domain is about 240 $\mu$A at 1.5 K. **d**, The magnetometry image obtained after cooling under zero field and zero current shows no flux contrast. **e** and **f**, Magnetometry image and current density $J$, respectively, under a constant current bias $I_b = 100$ $\mu$A reached from zero after the virgin cooling. The current density is obtained from the current flux image by a fast Fourier transform. The grayscale is the normalized intensity of $J$ and the streamlines trace the current flow (same below). **g** and **h**, Magnetometry image and current density, respectively, under $I_b = 200$ $\mu$A. Note that the flux signal in **g** changes sign from **e** with the same sample bias direction. **i**, Remanent flux signal after reaching $I_p = 200$ $\mu$A and then returning to $I_b = 0$ $\mu$A. **j** and **k**, Magnetometry image and current density, respectively, under $I_b = -200$ $\mu$A. **l**, Remanent flux signal after reaching $I_p = -200$ $\mu$A and then returning to $I_b = 0$ $\mu$A. All the scale bars are 20 $\mu$m. The remanent flux corresponds to an anomalous supercurrent distribution $J_A$ with a large magnitude on the edge.

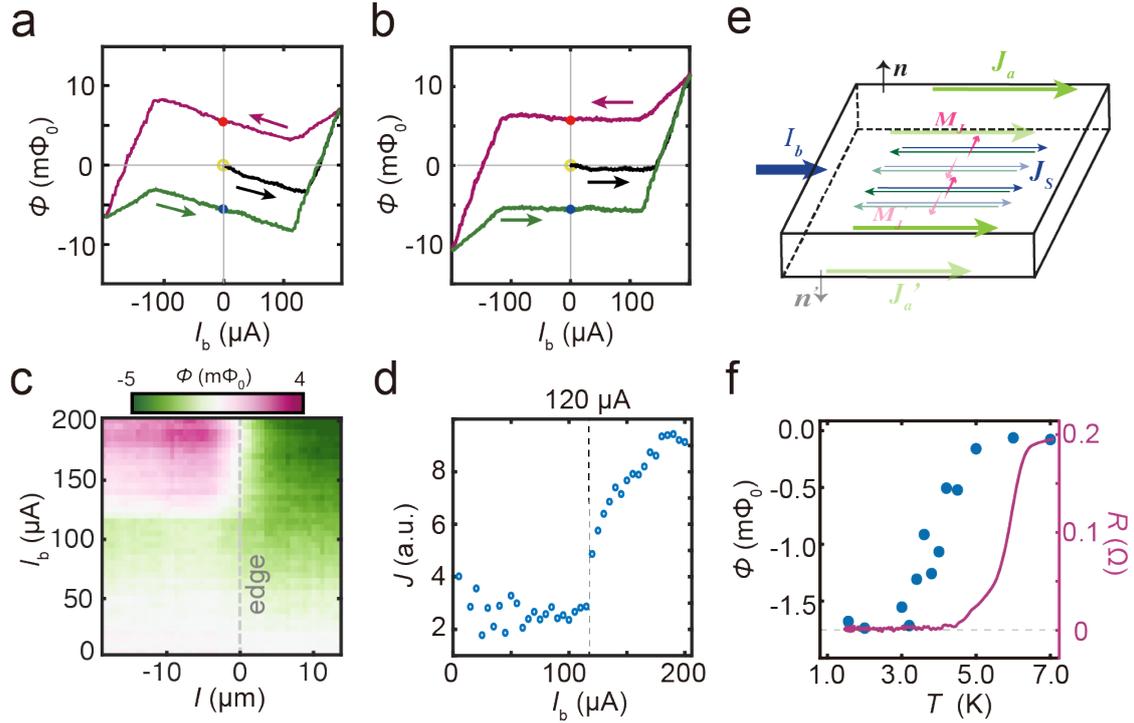

**Figure 3. Hysteresis of remanent flux and anomalous edge current. a**, Flux hysteresis loop as a function of $I_b$ obtained at a fixed location (Fig. 2d, yellow mark). The loop starts from a virgin cool (yellow circle at the origin), up sweeps to 200 μA (black points), down sweeps to −200 μA (magenta points) and then up sweeps to 200 μA (green points) to complete the loop. **b**, Flux hysteresis loop after subtracting the direct flux contribution of the bias current from that in **a**. **c**, The flux line cuts across the edge as labeled in Fig. 2e obtained under increasing $I_b$'s after a zero-field cool (Fig. S3). **d**, Peak supercurrent density on the same edge and extracted from the same flux images as in **c**. There is a sudden jump in edge current density at $I_b = 120$ μA. **e**, Illustration of the vectorial product relation between supercurrent and magnetization of a Dirac system with finite thickness. The bias induces a supercurrent distribution $J_s$ (blue arrows) flowing in the same direction in the top and bottom surfaces. The supercurrent leads to a surface magnetization $M_J$ (Magenta arrows) following $M_J \propto n \times J_S$, where $n$ is the surface normal. Magnetization on the bottom surface $M_J'$ is opposite from $M_J$. Due to the spin-galvanic (or magnetoelectric) effect, $M_J$ induces an anomalous current $J_a \propto n \times M_J$. $J_A$ in the bulk (dark green arrows) are opposite from $J_S$ while those on the edge (light green arrows) flow in the same direction. $J_a$ in the top and bottom surfaces are equal and in the same direction. After removing the bias, the remanent flux is largely due to the $J_a$, which is proportional to the remanent $M_J$ in each surface. **f**, Temperature dependence of resistance (magenta line) and the magnetization extracted from the images of Fig. S5.

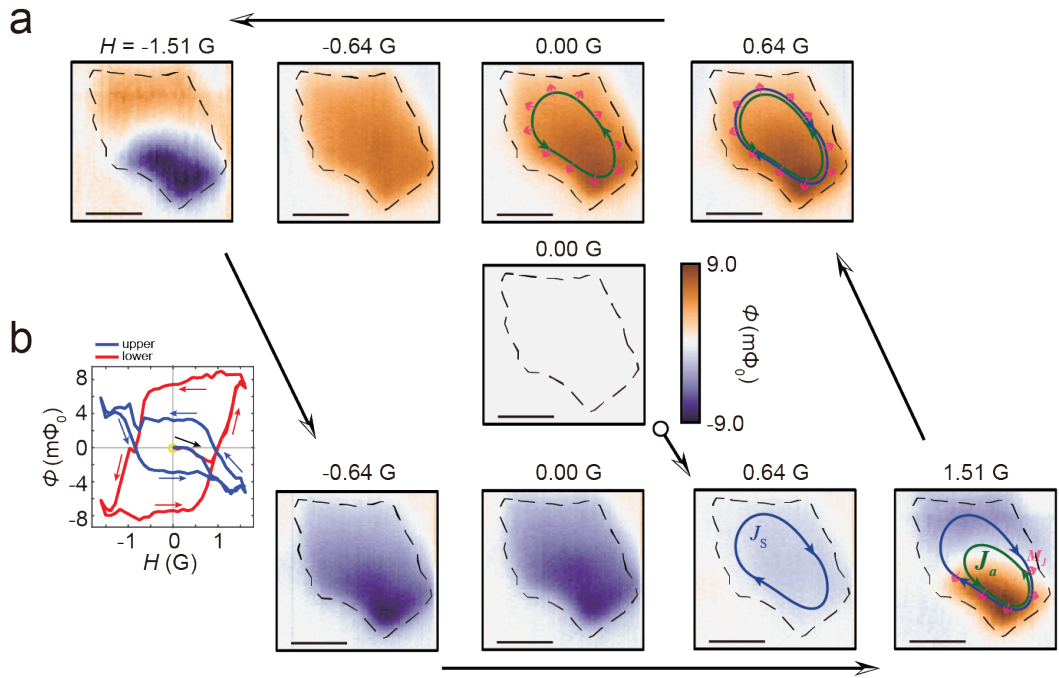

**Figure 4. Hysteretic switching by Meissner current. a**, Representative magnetometry images of the sample at different out-of-plane magnetic fields $H$ as indicated above the panels. The sequence starts with a virgin cool under zero field and zero bias current (the middle panel). The maximum $H = 1.51$ G applied during the sequence is much smaller than the lower critical field of Fe(Se,Te). As a result, the direct effect of $H$ is to induce a Meissner current circulating around the sample (blue circles). Scale bars are 20 $\mu$m. **b**, Magnetic flux of the lower (red) and upper (blue) domains as a function of $H$ extracted from magnetometry images (full set shown in Fig. S6). The sequence starts at the yellow circle.